# pMatlab Parallel Matlab Library

Nadya Travinin and Jeremy Kepner
{nt, kepner}@ll.mit.edu
MIT Lincoln Laboratory[†], 244 Wood Street, Lexington, MA 02420

**Abstract**

MATLAB® has emerged as one of the languages most commonly used by scientists and engineers for technical computing, with ~1,000,000 users worldwide. The primary benefits of MATLAB are reduced code development time via high levels of abstractions (e.g. first class multi-dimensional arrays and thousands of built in functions), interpretive, interactive programming, and powerful mathematical graphics. The compute intensive nature of technical computing means that many MATLAB users have codes that can significantly benefit from the increased performance offered by parallel computing. pMatlab (www.ll.mit.edu/pMatlab) provides this capability by implementing Parallel Global Array Semantics (PGAS) using standard operator overloading techniques. The core data structure in pMatlab is a distributed numerical array whose distribution onto multiple processors is specified with a "map" construct. Communication operations between distributed arrays are abstracted away from the user and pMatlab transparently supports redistribution between any block-cyclic-overlapped distributions up to four dimensions. pMatlab is built on top of the MatlabMPI communication library (www.ll.mit.edu/MatlabMPI) and runs on any combination of heterogeneous systems that support MATLAB, which includes Windows, Linux, MacOSX, and SunOS. This paper describes the overall design and architecture of the pMatlab implementation. Performance is validated by implementing the HPC Challenge benchmark suite and comparing pMatlab performance with the equivalent C+MPI codes. These results indicate that pMatlab can often achieve comparable performance to C+MPI at usually $1/10^{th}$ the code size. Finally, we present implementation data collected from a sample of 10 real pMatlab applications drawn from the ~100 users at MIT Lincoln Laboratory. These data indicate that users are typically able to go from a serial code to a well performing pMatlab code in about 3 hours while changing less than 1% of their code.

## 1. Introduction

MATLAB® has emerged as one of the predominant languages of technical computing. Its popularity for data analysis, simulation, and modeling is largely due to the expressiveness of the language, which approaches that of written mathematics. Additionally, MATLAB provides its users with powerful graphics that allow visualization of complex multi-dimensional datasets. The users of MATLAB tend to be engineers and scientists. High-level languages allow them to concentrate on their core competency and spend less effort on computer science-related implementation details. It is common for scientists and engineers to test the validity of data processing algorithms or physical simulations by employing larger data sets, higher resolution models, or a broader range of input parameters. This need for greater fidelity causes the execution times to reach hours or even days. Thus, a parallel capability that provides good speed

---

MATLAB® is a registered trademark of the MathWorks. Reference to commercial products, trade names, trademarks or manufacturer does not constitute or imply endorsement up without sacrificing the ease of programming is highly beneficial. pMatlab seeks to provide this capability by implementing standard Parallel Global Array Semantics (PGAS) (see Figure 1) using operator overloading techniques.

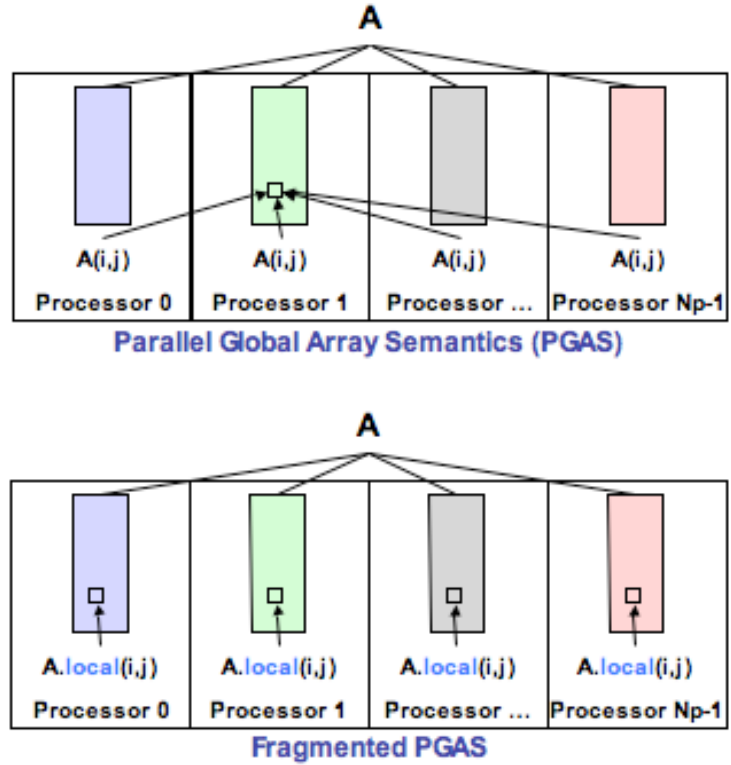

**Figure 1. Parallel Global Array Semantics (PGAS).** The top half of the figure illustrates pure PGAS. Matrix A is distributed among Np processors. Element $i,j$ is referenced on all the processors. In pure PGAS, the index $i,j$ is a global index which references the same element on all processors (on Processor 1 that element is local, on all other processors it is remote). The lower half of the figure illustrates fragmented PGAS. Here, each processor references element $i,j$ local to the processor, thus each processor references a different element in the global matrix. pMatlab supports both pure and fragmented PGAS.

The core data structures in pMatlab are distributed arrays and maps, which will be discussed in greater detail later in the paper. These data structures are illustrated in the pMatlab code fragment (see Figure 2) of the STREAM benchmark [McCalpin2005]. STREAM is a simple, embarrassingly parallel code that uses basic vector operations, such as scale and add, to measure main memory bandwidth. Distributed arrays allow the serial STREAM program to be quickly transformed into a parallel program by simply adding a "map" object to selected arrays. The map describes how the distributed array is to be broken up amongst multiple processors. Additionally, pMatlab also abstracts communication layer from the application developer. While writing a parallel MATLAB program with pMatlab, the user does not have to worry about parallel programming concepts such as deadlocks, barriers, and synchronization.

This paper describes the design, implementation and performance results of the pMatlab library used to create the constructs shown in Figure 2: the rest of this section highlights related work and different approaches to developing a parallel MATLAB capability. Section 2 addresses the details of the pMatlab design. Section 3 describes the implementation of the pMatlab library.

Section 4 presents the HPC Challenge implementations and benchmark results. Section 5 presents results from real applications. Section 6 presents our conclusions.

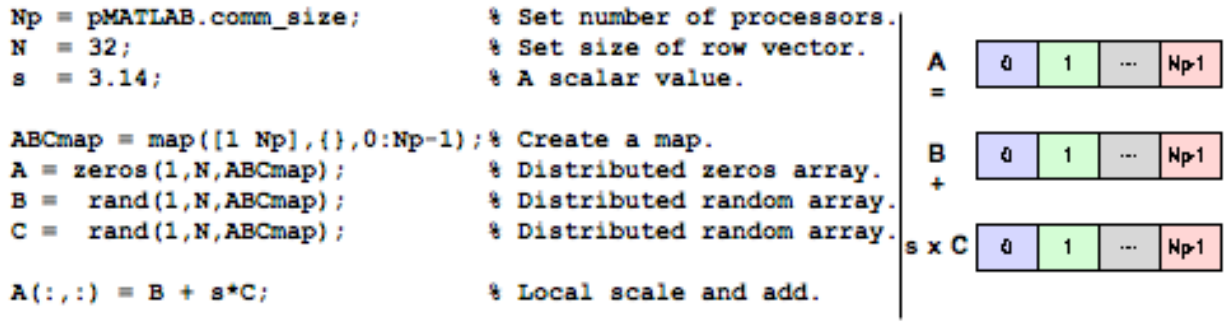

**Figure 2. STREAM Benchmark Code Highlights.** The first three lines set the various constants required by the program such as the number of processors and the size of the row vector. The next line creates a map, which will cause the $2^{nd}$ dimension of a distributed array to be broken up equally among all the processors. The next three lines use this map to create three row vectors. The last line performs the basic STREAM triad arithmetic operations in parallel. No communication is required in this example because A, B and C are all mapped the same.

## 1.1 Related Work

Parallel MATLAB has been an active area of research for a number of years and many different approaches have been developed (see [Choy2003] for a comprehensive survey). These different approaches can be roughly divided into three categories: message passing, client/server and global arrays.

The message passing approach [Kepner2004a, CMTM] requires the user to explicitly send messages within the code. These approaches often implement a variant of the Message Passing Interface (MPI) standard [MPI]. Message passing allows any processor to directly communicate with any other processor and provides the minimum required functionality to implement any parallel program. Users that are already familiar with MPI find these approaches powerful, but the learning curve is steep for the typical user because explicit message passing approaches significantly lower the level of abstraction and require users to deal directly with deadlocks, barriers, and other low level parallel programming constructs. In addition, the impact on code size is significant. Serial programs converted to parallel programs with MPI typically increase in size by 25% to 50%; in contrast, OpenMP and PGAS approaches typically only increase the code size by ~5% [Funk2005]. In spite of these difficulties, a message passing capability is a requirement for both the client/server and global arrays approaches. Furthermore, message passing is often the most efficient way to implement a program and there are certain programs with complex communication patterns that can only be implemented with direct message passing. Thus, any complete parallel solution must provide a mechanism for accessing the underlying messaging layer. Among the available MATLAB message passing implementations MatlabMPI [Kepner2004a] is currently the most popular implementation with thousands of users worldwide (see Section 3.4 for a more detailed discussion on MatlabMPI). More recently, the incorporation of MPI into The MathWorks' Distributed Computing Toolbox (DCT) [Dean2005] makes message passing available to a much broader range of users.

Client/server approaches [Choy2005, Morrow1998, RTExpress, Dean2005] use MATLAB as the user's front-end to a distributed library. For example, Star-P [Choy2005] keeps the distributed arrays on a parallel server, which calls the necessary routines from parallel libraries such as ScaLAPACK and FFTW. These approaches often provide the best performance once the data are transferred to the server. However, these approaches are limited to those functions that have been specifically linked to a parallel library and require the users to install additional libraries. Another potential disadvantage to these solutions is that backend libraries often require specific data distributions for their algorithms. For example, ScaLAPACK requires that the arrays be distributed in a 2D block-cyclic distribution (see Section 3.2, Maps). If the parallel MATLAB library does not support this type of distribution, extra communication overhead is incurred when redistributing the data for submission to the ScaLAPACK routine. We have included DCT in this category, although in this instance the back-end server is MATLAB running on each processor and the user is responsible for breaking up the calculation into embarrassingly parallel tasks that can be independently scheduled onto the workers.

pMatlab falls into the third category, the global arrays approach. Star-P [Choy2005] and Falcon [Falcon] also fall into this category. These approaches provide a mechanism for creating arrays, which are distributed across multiple processors. Global arrays have a long history in other languages, for example Fortran [Koelbel1994, Numrich1998] and C [El-Ghazawi2005], as well as in many C++ libraries such as POOMA [Cummings1998], GA Toolkit [Nieplocha2002], PVL [Lebak2005] and Parallel VSIPL++ [Lebak2005]. The global array approach allows the user to view a distributed object as a single entity, as opposed to multiple pieces as is the case with message passing. This approach allows operation on the array as a whole or on local parts of the array. Additionally, these libraries are compatible with MPI and are amenable to hybrid shared/distributed memory implementations. Parallel VSIPL++ is implemented for C++. The GA toolkit is implemented for a number of languages including Fortran, C, and C++.

pMatlab supports both pure PGAS and fragmented PGAS programming models (see Figure 1). The pure PGAS model presents an entirely global view of a distributed array. Specifically, once created with an appropriate map object, distributed arrays are treated the same as non-distributed ones. When using this programming model, the user never accesses the local part of the array and all operations (such as matrix multiplies, FFTs, convolutions, etc.) are performed on the global structure. The benefits of pure PGAS are ease of programming and the highest level of abstraction. The drawbacks include the need to implement parallel versions of serial operations and library performance overhead.

Fragmented PGAS maintains a high level of abstraction but allows access to local parts of the arrays. Specifically, a distributed array is created in the same manner as in pure PGAS, however, the operations can be performed on just the local part of the array. Later, the global structure can be updated with locally computed results. This allows greater flexibility. Additionally, this approach does not require function coverage or implementation of parallel versions of all existing serial functions. Furthermore, fragmented PGAS programs often achieve better performance by eliminating the library overhead on local computations.

pMatlab is a unique parallel MATLAB implementation for a number of reasons. pMatlab supports both pure and fragmented PGAS programming models, and allows combining PGAS with direct message passing for optimized performance. While pMatlab does use message passing in the library routines, a typical user does not have to explicitly incorporate messages into the code. pMatlab supports embarrassingly parallel computation, but is not limited to it. pMatlab does not link in any external libraries, nor does it compile the language into an executable. Our library is implemented entirely in MATLAB. This significantly reduces the size of the library and has allowed pMatlab to become the most complete implementation of PGAS available in any language.

## 2. pMatlab Interface and Architecture Design

The primary challenge in implementing a parallel computation library is how to balance the conflicting goals of ease of use, high performance, and ease of implementation. With respect to pMatlab, we have specifically defined each of these goals in a measurable way (see Table 1). The performance metrics are typical of those used throughout the high performance computing community and primarily look at the computation and memory overhead of programs written with pMatlab relative to serial programs written using MATLAB and parallel programs written using C with MPI. The metrics for ease of use and ease of implementation are derived from the software engineering community (see [Johnson2004, 2005] and [Kepner2004b] and papers therein) and look at code size, programmer effort, and required programmer expertise. These metrics are not perfect, but they are useful tools for measuring progress towards these goals. In the rest of this section we will discuss the particular choices made in pMatlab to satisfy these goals.

**Table 1: pMatlab Design Goals.** Metrics were defined for each of the high level pMatlab design goals: ease of use, performance, and ease of implementation. These metrics led to specific approaches for addressing the goals in a measurable way.

| Goal | Ease of use |
|---|---|
| Metrics | -Time for a user to produce a well performing parallel code from a serial code.<br>-Fraction of serial code that had to be modified.<br>-Expertise required to achieve good performance. |
| Approach | -Separate functional coding from mapping onto a parallel architecture.<br>-Abstract message passing away from the user.<br>-Ensure that simple (embarrassingly) parallel programs are simple to express.<br>-Provide a simple mechanism for globally turning pMatlab constructs on and off.<br>-Ensure backward compatibility with serial MATLAB.<br>-Provide a well-defined and repeatable process for migrating from serial to parallel code. |
| **Goal** | **High Performance** |
| Metrics | -Execution time and memory overhead as compared to serial MATLAB, the underlying MatlabMPI communication library and C+MPI benchmarks. |
| Approach | -Use underlying serial MATLAB routines wherever possible (even if it means slightly larger user code).<br>-Minimize the use of overloaded functions whose performance depends upon how distributed arrays are mapped.<br>-Provide a simple mechanism for using lower level communication when necessary. |
| **Goal** | **Ease of implementation** |
| Metrics | -Time to implement a well performing parallel library.<br>-Size of library code.<br>-Number of objects.<br>-Number of overloaded functions.<br>-Functional and performance test coverage. |
| Approach | -Utilize a layered design that separates math and communication.<br>-Leverage well-understood PGAS and data redistribution constructs.<br>-Minimize the use of overloaded functions.<br>-Develop a "pure" MATLAB implementation to minimize code size and maximize portability. |

## 2.1 Ease of use

The first step in writing a parallel program is to start with a functionally correct serial program. The conversion from serial to parallel requires the user to add new constructs to their code. pMatlab adopts a separation-of-concerns approach to this process which seeks to make functional programming and mapping a program to a parallel architecture orthogonal. A serial program is made parallel by adding maps to arrays. Maps only contain information about how an array is broken up onto multiple processors and the addition of a map *should not* change the functional correctness of a program. A map (see Figure 3) is composed of a grid specifying how each dimension is partitioned, a distribution that selects either a block, cyclic or block-cyclic partitioning, and a list of processors that defines which processors actually hold the data.

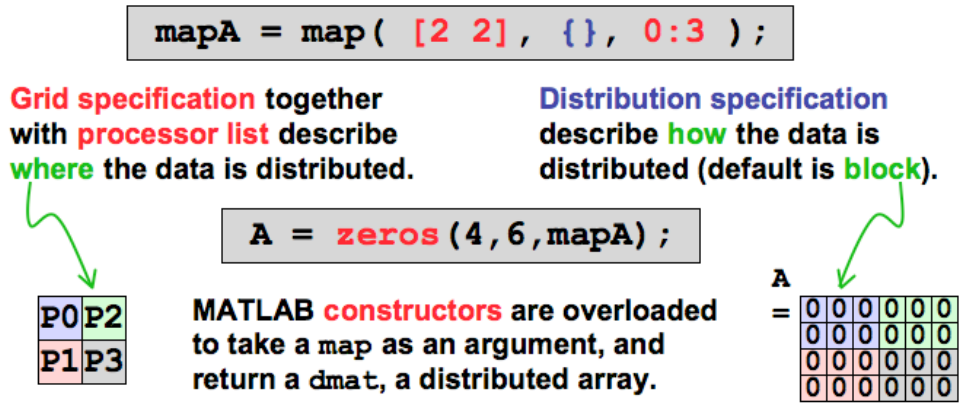

**Figure 3. Anatomy of a Map.** A map for a numerical array is an assignment of blocks of data to processing elements. It consists of a grid specification (in this case a 2 x 2 arrangement), a distribution (in this case {} implies that the default block distribution should be used), and a processor list (in this case the array is mapped to processors 0, 1, 2, and 3).

The next step in writing a parallel program is implementing communications. Perhaps the largest benefit of PGAS is the ability to abstract complex message passing away from the user. More specifically, redistribution between any two distributed arrays in pMatlab is accomplished with the "=" operator. In the STREAM benchmark example (see Figure 2) the "=" operator was used in the statement: `A(:,:) = B + s*C`, but since the arrays A, B and C all have the same map, no communication was required. The overloaded "=" operator in pMatlab figures this out and correctly performs a simple assignment of the local data on the right hand side to the local data on left hand sized. A more complex example is the HPC Challenge FFT benchmark (see Figure 4). This benchmark computes the Fast Fourier Transform of a large 1D vector. The standard parallel algorithm for this benchmark is to transform the 1D vector into a row distributed matrix, FFT the rows of the matrix, multiply by a set of weights, redistribute into a column distributed matrix, and FFT the columns. A key step in the process is the redistribution which is performed by the statement: `Z(:,:) = X`, which determines and executes the $Np^2$ messages that need to be sent to complete this operation. (Note: pMatlab maps also allow this operation to be performed using a pipeline by using different processor sets in the maps; this capability is discussed further in section 3.2.)

```
1. Np = pMATLAB.comm_size;   % Set number of processors.
2. P = 2^10;   Q = 2^10;     % Set dimensions of array.

3. Xmap = map([Np 1],{},0:Np-1);   % Row map.
4. Zmap = map([1 Np],{},0:Np-1);   % Column map.
% Create complex global arrays X and Z for FFT.
5. X = complex(rand(P,Q,Xmap),rand(P,Q,Xmap));
6. Z = complex(zeros(P,Q,Zmap));

7. X = fft(X,[],2);          % FFT rows.
8. X = X.* W;.                % Multiply by weights.
9. Z(:,:) = X;               % Redistribute data.
10.Z = fft(Z,[]1);           % FFT columns.
```

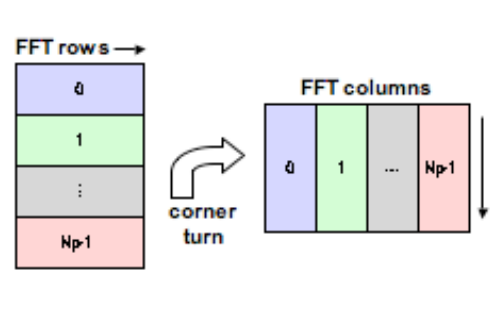

**Figure 4. FFT Benchmark Code Highlights.** The first two lines set the various constants required by the program such as the number of processors and the size of the matrix. The next two lines create two map objects for breaking the matrix up into rows and into columns. The next two lines use the maps to create two matrices. The next four lines FFT the rows, multiply by a set of local pre-computed weights, redistribute the data (using the "=" operator) into the matrix broken up by columns, and then perform the FFT on the columns.

PGAS enables complex data movements to be expressed compactly without making parallelism a burden to code. For example, removing the maps from either the STREAM or FFT examples returns the program to a valid serial program that simply used standard built-in operations. This is a direct a result of the orthogonality of mapping and functionality, and allows the pMatlab library to be "turned off" by simply setting all the maps equal to the scalar value of 1. This feature exploits a side effect of MATLAB constructors (e.g. zeros and rand), which ignore a trailing argument equal to 1. This ability to turn the library on and off is a key debugging feature and allows users to determine whether the bugs are from problems in their serial code or due to their use of pMatlab constructs.

All of these steps: making the code parallel, managing the communication and debugging, need to be directly supported in the library. Our experience with many pMatlab users has resulted in a standardized and repeatable process (see Figure 5) for quickly going from a serial code to a well-performing parallel code. This process is very important, as the natural tendency of new pMatlab users is to add parallel functions and immediately attempt to run large problems on a large number of processors.

The four step process begins by adding distributed matrices to the serial program, but then assigning all the maps to a value of 1 and verifying the program with Np=1 on the local machine. The second step is to turn the maps on and to run the program again with Np=1, which will verify that the pMatlab overloading constructs are working properly. It is also important to look at the relative performance of the first and second steps, as this will indicate if any unforeseen overheads are incurred by using the pMatlab constructs. The third step is to run with Np>1 on the local machine, which will verify that the pMatlab communication constructs are working properly. The fourth and final step is to run with Np>1 on multiple machines, which validates that the remote communication is working properly. Only after these four steps have been performed is it worthwhile to attempt to run large problems on many processors. In addition, it is important to always debug problems at the lowest numbered step.

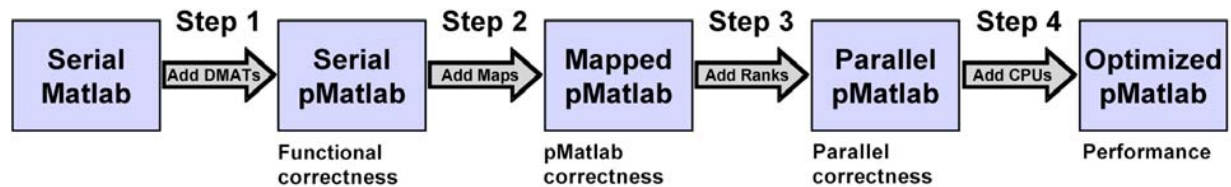

**Figure 5. The four step serial to parallel process.** Step 1 adds distributed matrices to the serial program, then assigns all the maps a value of 1 and runs with Np=1 on the local machine. Step 2 turns the maps on and runs the program again with Np=1. Step 3 runs with Np>1 on the local machine. Step 4 runs with Np>1 on multiple machines. Debugging should always be performed at the lowest numbered step where a problem occurs.

## 2.2 High Performance

The primary goal of using a parallel computer is to improve runtime performance. The first step in achieving high performance is to minimize the overhead of using pMatlab constructs as compared to their serial equivalents. The previous examples (Figures 2 and 4) show the ideal "pure" PGAS case when all the required functions have been overloaded to work well with pMatlab distributed arrays. It is impractical (and unnecessary) to provide optimized implementations of the approximately 8,000 built-in functions for every combination of array distributions. Instead, we adopt a coding style that uses some fragmented PGAS constructs (see Section 1.1). This style is less elegant but provides strict guarantees on performance. More specifically, distributed arrays are used as little as possible and only when interprocessor communication is required.

Figure 6 shows examples of the STREAM and FFT benchmarks written using fragmented PGAS constructs that minimize the use of overloaded functions by employing the `local` and `put_local` functions (see section 3.5). The `local` function extracts the local part of the distributed array and returns a regular MATLAB array that will work with any serial MATLAB function. The `put_local` function replaces the local part of a distributed array with a regular serial MATLAB array. Thus, in the STREAM and FFT examples the key expression: `Alocal = Blocal + s*Clocal`, and `fft(local(X),[],2)` are guaranteed to have the same performance as the equivalent serial function calls and eliminate the need for pMatlab to overload +, * and fft. In addition to providing a local performance guarantee this style of coding minimizes the potential for "accidental" communication which is easy to do with the "=" operator. This style of coding has proven to be very effective and most users are able to adapt their code to this style with minimum effort. In support of this style, the pMatlab library also provides serial equivalents of the `local` and `put_local` functions so that the code will still work if parallel arrays are turned off.

The power of PGAS is its ability to hide underlying communication from the user and eliminate the need for writing lengthy and complex message passing code. Unfortunately, PGAS constructs are not appropriate for all circumstances. There are communication patterns that simply would be more efficient if direct message passing can be employed. Thus, it is important to have mechanisms that allow PGAS and the underlying communication constructs to interact easily. pMatlab provides this ability by allowing the user to directly access the underlying MatlabMPI library and its data structures. At any time in the program the user, if s/he so desires, can choose to send messages directly with MatlabMPI. In fact, we have found that PGAS and message passing work very well together since the PGAS constructs can still be used to quickly figure out which data to send and where to send it.

Several of the HPC Challenge benchmarks fall into the class of codes that do best by allowing some use of direct message passing. In the case of the FFT code, we have used a special function called `transpose_grid` (see Figure 6) that directly uses MatlabMPI messaging to optimally perform the all-to-all communication for going from a row distributed matrix to a column distributed matrix. This function is able to use memory more efficiently and to optimize the order in which messages are sent and received. The RandomAccess benchmark (see section 4.3) requires that all processors are able to randomly communicate with all other processors and is a more explicit example of using messaging and PGAS together. The HPL Top500 benchmark (see section 4.4) requires that one processor be able to broadcast to a subset of all the other processors, which is also most easily dealt with using direct message passing.

```
Optimized STREAM code                      Optimized FFT code
Np = pMATLAB.comm_size;                    Np = pMATLAB.comm_size;
N = 32;                                    P = 2^10;   Q = 2^10;
s = 3.14;

ABCmap = map([1 Np],{},0:Np-1);            Xmap = map([Np 1],{},0:Np-1);
Alocal = local(zeros(1,N,ABCmap));         X = complex(rand(P,Q,Xmap),rand(P,Q,Xmap));
Blocal = local(rand(1,N,ABCmap));
Clocal = local(rand(1,N,ABCmap));          X = put_local(X, fft(local(X),[],2) .* Wlocal );
                                           Z = transpose_grid(X);
Alocal = Blocal + s*Clocal;                Z = put_local(Z, fft(local(Z),[],1) );
```

**Figure 6. Optimized STREAM and FFT Code Highlights.** Programs have been rewritten to minimize the number of overloaded functions required by using the `local` and `put_local` functions. These programs are guaranteed to provide the same local performance as their serial equivalents.

## 2.3 Ease of implementation

The ease of use and high performance goals are well understood by the HPC community. Unfortunately, implementing these goals in a middleware library often proves to be quite costly. A typical PGAS C++ library can be 50,000 lines of code and requires several programmers years to implement. pMatlab has adopted several strategies to reduce implementation costs. The common theme among these strategies is finding the minimum set of features that will still allow users to write well performing programs.

One of the key choices in implementing a PGAS library is which data distributions (see section 3.2) to support? At one extreme it can be argued that most users are satisfied by 1D block distributions. At the other extreme, one can find applications that require truly arbitrary distributions of array indices to processors. pMatlab has chosen to support all 4D block-cyclic distributions with overlap because the problem of redistribution between any two such distributions (see section 3.3) has been solved a number of times by different parallel computing technologies.

The pMatlab "=" operator supports data redistribution between arrays. The next question is what other functions to support and for which distributions? Table 2 shows an enumeration of different levels of PGAS support. The ability to work with the local part of a distributed array and its indices has also been demonstrated repeatedly. The big challenge is overloading all mathematical functions in a library to work well with every combination of input distributions. As discussed in section 2.2, this capability is extremely difficult to implement and is not entirely necessary if users are willing to tolerate the slightly less elegant coding style associated with fragmented PGAS. Thus, pMatlab provides a rich set of data distributions, but a relatively modest set of overloaded functions, which are mainly focused on array construction functions, array index support functions, and the various element-wise operations (+,-,.*,./, …).

The final implementation choice was to implement pMatlab purely in MATLAB without relying on binding to other languages. This has minimized code size and maximized portability. For example, pMatlab is the most complete implementation of PGAS, but it is only about 3,000 lines of code and has introduced only two new objects (maps and distributed arrays). pMatlab also runs on any combination of heterogeneous systems that support MATLAB, which includes Windows, Linux, MacOSX, and SunOS.

**Table 2: Lebak Levels.** Levels of parallel support for data and functions. Note: Support for data distribution is assumed to include support for overlap in any distributed dimension. Data4/Op1 has been successfully implemented many times. Data1/Op2 may be possible but has not yet been demonstrated.

| Data Level | Description of Support |
| --- | --- |
| Data0 | Distribution of data is not supported [not a parallel implementation] |
| Data1 | One dimension of data may be block distributed |
| Data2 | Two dimensions of data may be block distributed |
| Data3 | Any and all dimensions of data may be block distributed |
| Data4 | Any and all dimensions of data may be block or cyclicly distributed. |

| Operations Level | Description of Support |
| --- | --- |
| Op0 | No distributed operations supported [not a parallel implementation] |
| Op1 | Distributed assignment, get, and put operations, and support for obtaining data and indices of local data from a distributed object. |
| Op2 | Distributed operation support (the implementation must state which operations those are) |

### 3. pMatlab Implementation

This section discusses the implementation details of the pMatlab library. The library is designed and implemented at MIT Lincoln Laboratory and builds upon concepts from the Parallel Vector Library (PVL) [Lebak2005], Star-P [Choy2005], and uses MatlabMPI [Kepner2004a] as the communication layer. Figure 7 illustrates the layered architecture of the parallel library. In the layered architecture, the pMatlab library implements distributed constructs, such as distributed matrices and higher dimensional arrays. In addition, pMatlab provides parallel implementations of a select number of functions such as redistribution, Fast Fourier Transform (FFT), and matrix multiplication. However, it is usually simpler for a user to create a parallel implementation of a function focused on his/her particular data sizes and data distributions of interests, than to provide generic parallel implementations of functions which give good performance for all data distributions and data sizes.

The pMatlab library uses the parallelism through polymorphism approach as discussed in [Choy2005]. Monomorphic languages require that each variable is of only one type; on the other hand in polymorphic languages variables can be of different types and polymorphic functions can operate on different types of variables [Cardelli1985]. The concept of polymorphism is inherent in the MATLAB language – variable types do not have to be defined, variable types can change during the execution of the program, and many functions operate on a variety of data types such as double, single, complex, etc.

In pMatlab, as in Star-P, this concept is taken one step further. The polymorphism is exploited by introducing the map object. Map objects belong to a pMatlab class *map* and are created by specifying the grid description, distribution description, and the processor list as discussed in section 2.1 (see Figure 3). The map object can then be passed to a MATLAB constructor, such as `rand`, `zeros`, or `ones`. The constructors are overloaded and when a `map` object is passed into a constructor, the library creates a variable of type `dmat`, or a distributed array. A PITFALLS structure (see section 3.3) is created when each `dmat` object is constructed. A PITFALLS is a mathematical representation of the data distribution information. pMatlab supports numerical arrays of up to four dimensions of different numerical data types and allows creation of distributed sparse matrices.

As discussed previously, a subset of functions, such as `plus`, `minus`, `fft`, `mtimes`, and all element-wise operations are overloaded to operate on `dmat` objects. When using a pure PGAS programming model and an overloaded function, the `dmat` object can be treated as a regular array. Functions that operate only on the local part of the `dmat` structure (element-wise operations) simply perform the operations requested on the `dmat.local` array, which is a standard MATLAB numerical type specified at array creation. Functions that require communication, such as redistribution (or `subsasgn` in MATLAB syntax) use MatlabMPI as the communication layer.

Let us return to the pMatlab FFT code in Figure 4. Lines 3 and 4 define two pMatlab map objects: Xmap and Zmap. The user defines maps to specify how and where the numerical arrays in the program are mapped. In this example all available processors are used (numbered sequentially from 0 to Np-1). Distributed arrays are created using the standard MATLAB array constructors: `zeros()`, `rand()`, and `ones()`. The outputs of the overloaded constructors are `dmat`s, or distributed arrays. Lines 5 and 6 in Figure 4 create two distributed complex matrices split up among Np processors. Xmap indicates that the matrix should be distributed row-wise with P/Np rows per processor, where as Zmap defines a column-wise distribution with Q/Np columns per processor. If a dimension is not evenly divisible by Np, pMatlab figures this out and shorts the last processor. Line 7 calls the overloaded FFT function on the distributed array X and returns the result into an array with same map as the input. Line 9 uses the overloaded "=" operator which performs an all-to-all communication which results in Z having the same data as X, while distributing this data according to the distribution defined in Zmap.

Since all functions supported in pMatlab are implemented in pure MATLAB, the pMatlab library maintains the portability of MatlabMPI. pMatlab can run anywhere MATLAB runs, given that there exists a common file system, a constraint imposed on pMatlab by MatlabMPI. A further benefit of the layered architecture of pMatlab is that any other communication library could be substituted for MatlabMPI, given that it implements the six basic MPI functions required by pMatlab (see section 3.4).

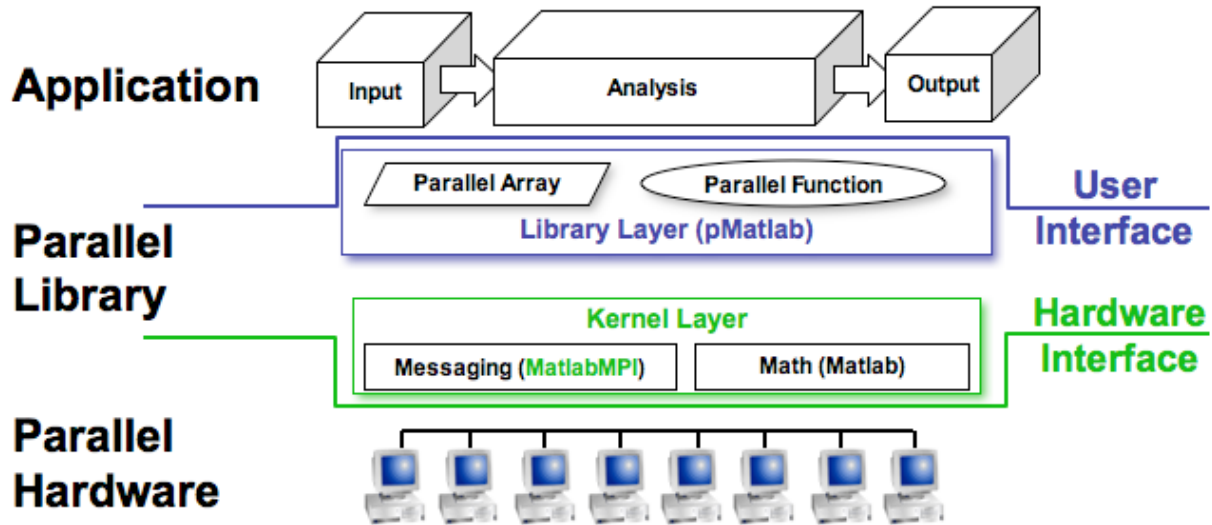

**Figure 7. Layered architecture.** The pMatlab library implements distributed constructs, such as vectors, matrices, and multi-dimensional arrays and parallel algorithms that operate on those constructs, such as redistribution, Fast Fourier Transform (FFT), and matrix multiplication.

### 3.1 pMatlab Execution

All pMatlab code resides within a generic code framework (see Figure 8) for initializing pMatlab (`pMatlab_Init`), determining the number of processors the program is being run on (`pMATLAB.comm_size`), determining the rank of the local processor (`pMATLAB.my_rank`), and finalizing the pMatlab library when the computation is complete (`pMatlab_Finalize`). pMatlab uses the Single Program Multiple Data (SPMD) execution model. The user runs a pMatlab program by calling the MatlabMPI `MPI_Run` command to launch and initialize the multiple instances of MATLAB required to run in parallel. Figure 8 shows an example RUN.m script using `MPI_Run` to launch four copies of the pFFT.m script.

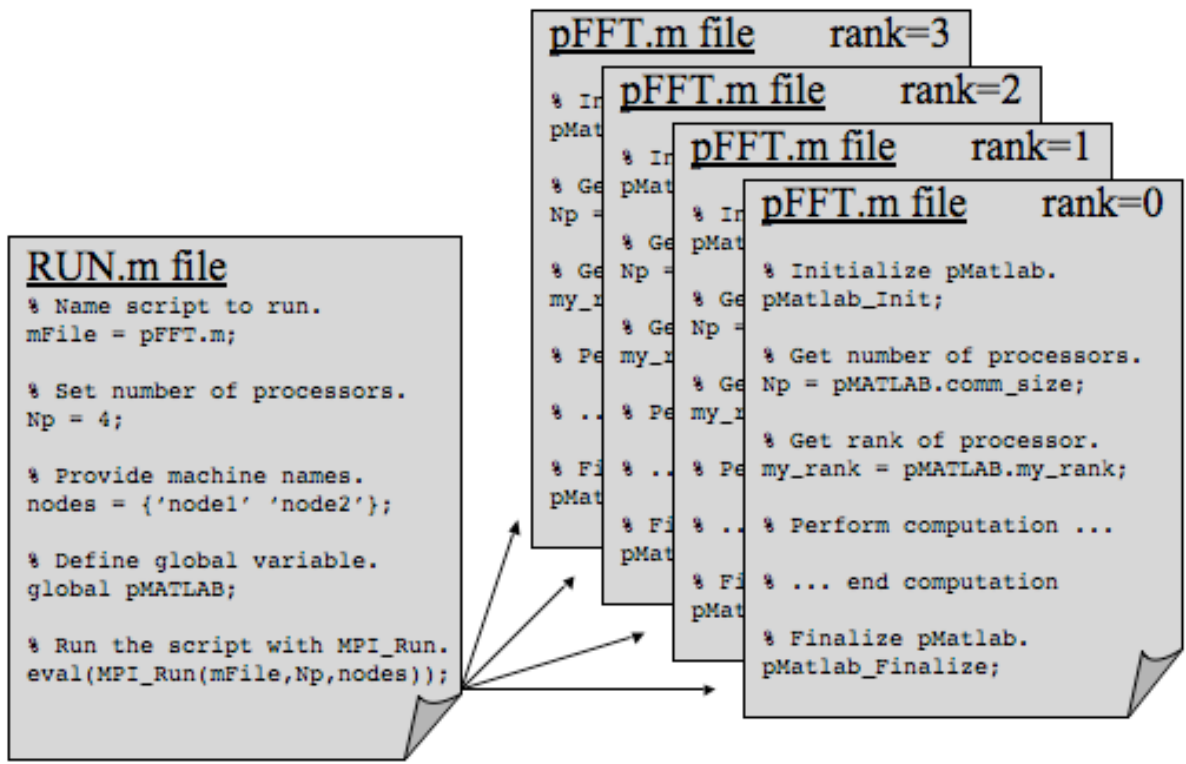

**Figure 8. pMatlab Execution Framework.** A pMatlab program (pFFT.m) is launched using the MPI_Run command shown in the RUN.m file, which sets the number of processors and the precise machines to run on. MPI_Run starts Np instances of Matlab each with a different rank. Within the pMatlab program the pMatlab environment is initialized and the number of processors and local rank can be obtained. The program is completed with the pMatlab_Finalize command.

### 3.2 Maps and Distributions

The concept of using maps to describe array distributions has a long history. The ideas for pMatlab maps are principally drawn from the High Performance Fortran (HPF) community [Loveman1993, Zosel1993], MIT Lincoln Laboratory Space-Time Adaptive Processing Library (STAPL) [DeLuca1997], and Parallel Vector Library (PVL) [Lebak2005]. A map for a numerical array defines how and where the array is distributed (Figure 3). PVL also supports task parallelism with explicit maps for modules of computation. pMatlab explicitly only supports data parallelism, however implicit task parallelism can be implemented through careful mapping of data arrays.

The pMatlab map construct is defined by three components: (1) grid description, (2) distribution description, and (3) processor list. The grid description together with the processor list describes where the data object is distributed, while the distribution describes how the object is distributed (see Figure 3). pMatlab supports any combination of block-cyclic distributions up to four dimensions. The API for defining these distributions is shown in Figure 9.

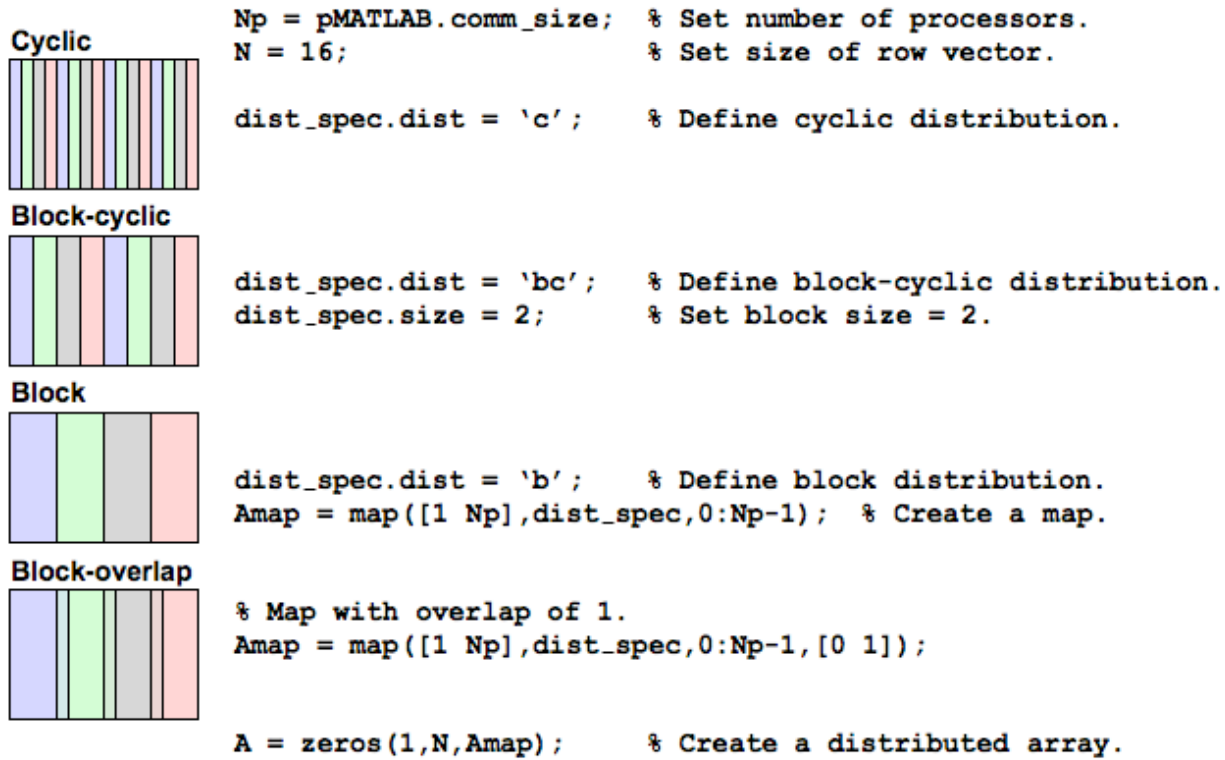

**Figure 9. Supported Distributions.**. Block distribution divides the object evenly among available processors. Cyclic distribution places a single element on each available processor and then repeats. Block-cyclic distributions places the specified number of elements on each available processor and then repeats.

Block distribution is the default distribution, which can be specified explicitly or by simply passing an empty distribution specification to the map constructor. Cyclic and block-cyclic distributions require the user to provide more information. Distributions can be defined for each dimension and each dimension could potentially have a different distribution scheme. Additionally, if only a single distribution is specified and the grid indicates that more than one dimension is distributed, that distribution is applied to each dimension.

Some applications, particularly image processing, require data overlap, or replicating rows or columns of data on neighboring processors. This capability is also supported through the map interface. If overlap is necessary, it is specified as an additional fourth argument. In Figure 9, the fourth argument indicates that there is 0 overlap between rows and 1 column overlap between columns. Overlap can be defined for any dimension and does not have to be the same across dimensions.

While maps introduce a new construct and potentially reduce the ease of programming, they have significant advantages over both message passing approaches and predefined limited distribution approaches. Specifically, pMatlab maps are scalable, allow optimal distributions for different algorithms, and support pipelining.

Maps are scalable in both the size of the data and the number of processors. Maps allow the user to separate the task of mapping the application from the task of writing the application. Different sets of maps do not require changes to be made to the application code. Specifically, the distribution of the data and the number of processors can be changed without making any changes to the algorithm. Separating mapping of the program from the functional programming is an important design approach in pMatlab (see Section 2.1).

Maps make it easy to specify different distributions to support *different algorithms*. Optimal or suggested distributions exist for many specific computations. For example, matrix multiply operations are most efficient on processor grids that are transposes of each other. Column and row wise FFT operations produce linear speed up if the dimension along which the array is broken up matches the dimension on which the FFT is performed (see Figure 4).

Maps also allow the user to set up *pipelines* in the computation, thus supporting implicit task parallelism. For example, pipelining is a common approach to hiding the latency of the all-to-all communication required in parallel FFT. The following slight change in the maps can be used to set up a pipeline where the first half of the processors perform the first part of the FFT and the second half perform the second part

```
Xmap = map([Np/2 1],{},[0    :Np/2-1]); % Row map on 1st set of cpus.
Zmap = map([1 Np/2],{},[Np/2:Np-1]);   % Column map on 2nd set of cpus.
```

When a processor encounters such a map, it first checks if it has any data to operate on. If the processor doesn't have any data it proceeds to the next line. In the case of the FFT with the above mappings, the first half of the processors (rank 0 to Np/2-1) will simply perform the row FFT, send data to the second set of processors, and skip the column FFT, and proceed to process the next set of data. Likewise, the second set of processors (ranks Np/2 to Np-1) will skip the row FFT, receive data from the first set of processors, and perform the column FFT.

### 3.3 Processor Indexed Tagged FAmiLy of Line Segments (PITFALLS)

Here we discuss an efficient and general technique for data redistribution. Such techniques are necessary in order to support PGAS. We chose to use PITFALLS [Ramaswamy1995], which is a mathematical representation of the data distribution. Additionally, [Ramaswamy1995] provides an algorithm for determining which pairs of processors need to communicate when redistribution is required and exactly what data needs to be sent.

A PITFALLS P is defined by the following tuple:

*P = (l, r, s, n, d, p)*

where
- *l* – starting index
- *r* – ending index
- *s* – stride between successive *l*'s
- *n* – number of equally spaced, equally sized blocks of elements per processor
- *d* – spacing between *l*'s of successive processor FALLS
- *p* – number of processors

The PITFALLS intersection algorithm is used to determine the necessary messages for redistribution. The algorithm can be applied to each dimension of the array, thus allowing efficient redistribution of arbitrary dimensional arrays. For a detailed discussion of the algorithm and its efficiency see [Ramaswamy1995]. (Note that the PITFALLS tuple can be derived in a trivial manner from the map definition.)

### 3.4 MatlabMPI

MatlabMPI [Kepner2004a] is a pure MATLAB implementation of the most basic MPI [MPI] functions. The functions required by pMatlab are listed in Table 3. The communication is done through file I/O (see Figure 10) through a common file system. The advantage of this approach is that the library is very small (~300 lines) and is highly portable. The price for this portability is that the while MatlabMPI performance is comparable to C+MPI for large messages, its latency for small messages is much higher (see Figure 11).

When designing pMatlab, it was important to ensure that the overhead incurred by the library did not significantly impact performance. From a library perspective, this means that the performance of the communication operations using the overloaded "=" operator should be as close as possible to the equivalent MatlabMPI code. Figure 12 shows the performance of an all-to-all operation using MatlabMPI, pMatlab "=" and the pMatlab `transpose_grid` function.

From an application perspective minimizing overhead means using algorithms that use fewer larger messages instead of many smaller messages. In Section 4 we will see that the relative performance of the HPC Challenge benchmarks can essentially be derived from the performance of the underlying MatlabMPI library. STREAM (no communication), FFT (all-to-all), and Top500 (broadcast) all fall into the large message regime and deliver reasonable performance. RandomAccess is designed to stress small messages and the relative performance of pMatlab is much worse. Fortunately, most real pMatlab programs tend to involve large messages.

**Table 3: Selected MPI functions provided by MatlabMPI.** pMatlab can be built on top of any communication library that implements these six functions.

| Function Name | Function Description |
|---|---|
| `MPI_Init` | Initializes MPI. |
| `MPI_Comm_size` | Gets the number of processors in a communication. |
| `MPI_Comm_rank` | Gets the rank of current processor within a communicator. |
| `MPI_Send` | Sends a message to a processor. |
| `MPI_Recv` | Receives a message from a processor. |
| `MPI_Finalize` | Finalizes MPI. |

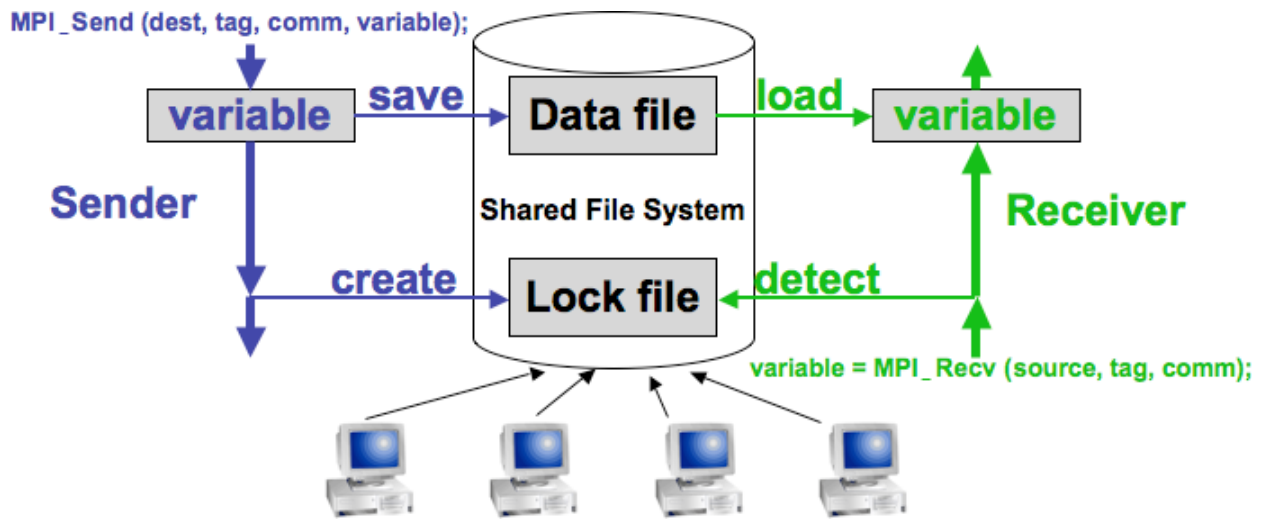

**Figure 10. MatlabMPI file I/O based communication.** MatlabMPI uses file I/O to implement point-to-point communication. The sender writes variables to a buffer file and then writes a lock file. The receiver waits until it sees the lock file, it then reads in the buffer file.

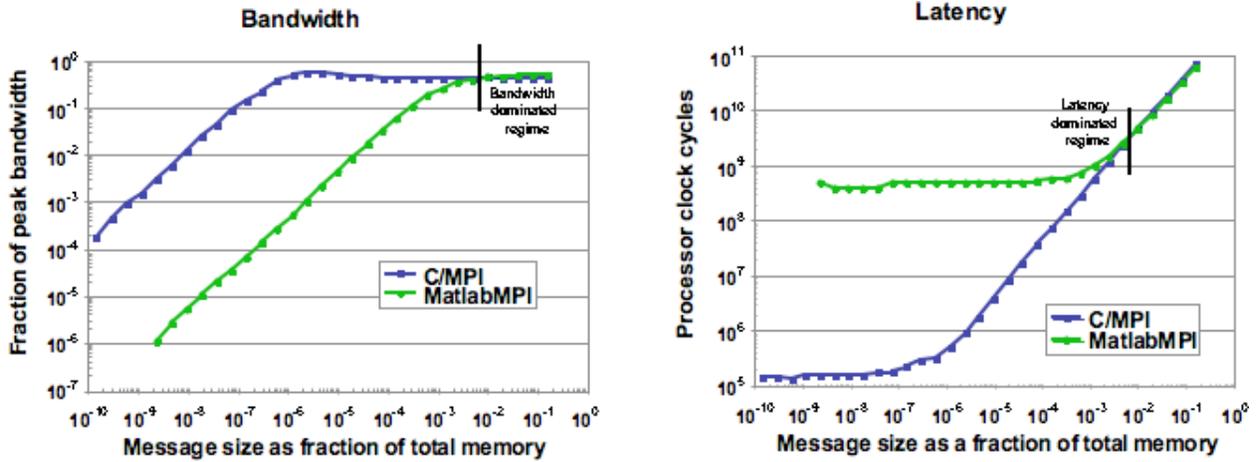

**Figure 11. MatlabMPI vs C+MPI.** Bandwidth and latency vs message size. Bandwidth is given as fraction of the peak underlying link bandwidth. Latency is given in terms of processor cycles. For large messages the performance is comparable. For small messages the latency of MatlabMPI is much higher.

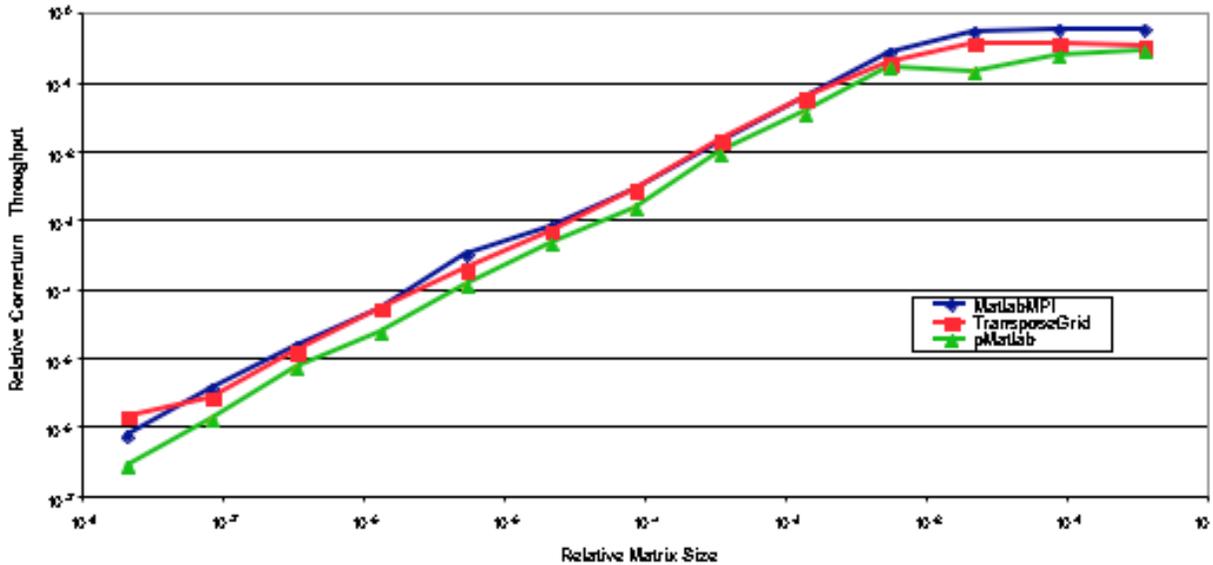

**Figure 12. MatlabMPI vs pMatlab.** Relative all-to-all performance for a pure MatlabMPI implementation, an A(:,:) = B implementation and a transpose grid implementation. The X-axis represents size of each matrix relative to node memory. The Y-axis represents throughput relative to peak bandwidth.

**3.5 pMatlab Parallel Support Functions**

Every PGAS implementation must provide a set of functions for managing and working with global arrays, which have no serial equivalents. The set of pMatlab parallel support functions is shown in Table 4. These functions allow the user to aggregate data onto one or many processors, determine which global indices are local to which processors, and get/put data from/to the local part of a distributed array. This set of functions is relatively small. To support the development process discussed in section 2.1, all of these functions have been overloaded to also work on serial MATLAB arrays so that the code will still work if the pMatlab maps have been turned off.

Table 4: pMatlab Parallel Support Functions.

| Function Name | Function Description |
|---|---|
| `synch` | synchronize the data in the distributed matrix. |
| `agg` | aggregates the parts of a distributed matrix on the leader processor. |
| `agg_all` | aggregates the parts of a distributed matrix on all processors in the communication scope |
| `global_block_range` | returns the ranges of global indices local to the current processor |
| `global_block_ranges` | returns the ranges of global indices for all processors in the map of distributed array D on all processors in communication scope |
| `global_ind` | returns the global indices local to the current processor |
| `global_inds` | returns the global indices for all processors in the map of distributed array D |
| `global_range` | returns the ranges of global indices local to the current processor |
| `global_ranges` | returns the ranges of global indices for all processors in the map of distributed array D |
| `local` | returns the local part of the distributed array |
| `put_local` | assigns new data to the local part of the distributed array |
| `grid` | returns the processor grid onto which the distributed array is mapped |
| `inmap` | checks if a processor is in the map |

## 4. HPC Challenge Benchmarks

In this section we focus on benchmark results to determine the limits of pMatlab performance. We are interested in looking at performance from a number of viewpoints. First, we are interested in the performance of pMatlab relative to serial MATLAB since this is what most users care about. Second, we are interested in the performance of pMatlab relative C+MPI as way of gauging the quality of the implementation and as a guide to future performance enhancements. We have chosen to use the HPC Challenge Benchmark suite [Luszczek2005] for this comparison (see Figure 13). HPC Challenge is designed to look at a range of computations that focus on different parts of the memory hierarchy. In addition, HPC Challenge computations are sufficiently well defined so that they can be implemented using a variety of programming models. We will first present the performance results and then discuss each of the benchmarks in more detail.

The four primary HPC Challenge benchmarks (STREAM, FFT, Top500 and RandomAccess) were implemented using pMatlab and run on a commodity cluster system (see Appendix A for a precise description of the hardware). Both the pMatlab and C+MPI reference implementation of the benchmarks were run on 1, 2, 4, 8, 16, 32, 64 and 128 processors. At each processor count the largest problem size was run that would fit in the main memory. The collected data measures the relative compute performance and memory overhead of pMatlab with respect to C+MPI (see Figure 14). In addition, we will also look at the relative code sizes of the benchmarks as an approximate measure of the complexity of the implementations. The relative memory, performance and code sizes are summarized in Table 5.

In general we see that the pMatlab implementations can run problems that are typically ½ the size of C+MPI implementation problem size. This is mostly due to the need to need to create temporary arrays when using high-level expressions. The pMatlab performance ranges from being comparable to the C+MPI code (FFT and STREAM), to somewhat slower (Top500), to a lot slower (RandomAccess). In contrast the pMatlab code is typically 3x to 40x smaller than the equivalent C+MPI code.

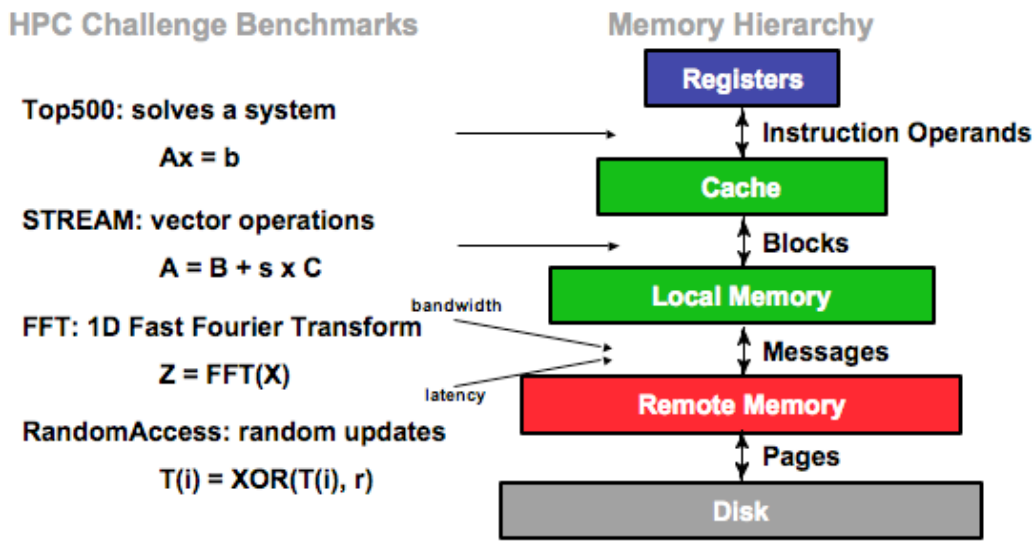

**Figure 13. HPC Challenge and the Memory Hierachy.** HPC Challenge benchmarks have been chosen to cover a range of memory access patterns and stress different parts of the memory hierarchy. Top500 performance is mostly dominated by local matrix multiply operations. RandomAccess is dominated by all-to-all communications of very small messages. FFT is also dominated by all-to-all communications, but for very large messages. STREAM requires no communication, is dominated by local vector operations, and stresses local processor to memory bandwidth.

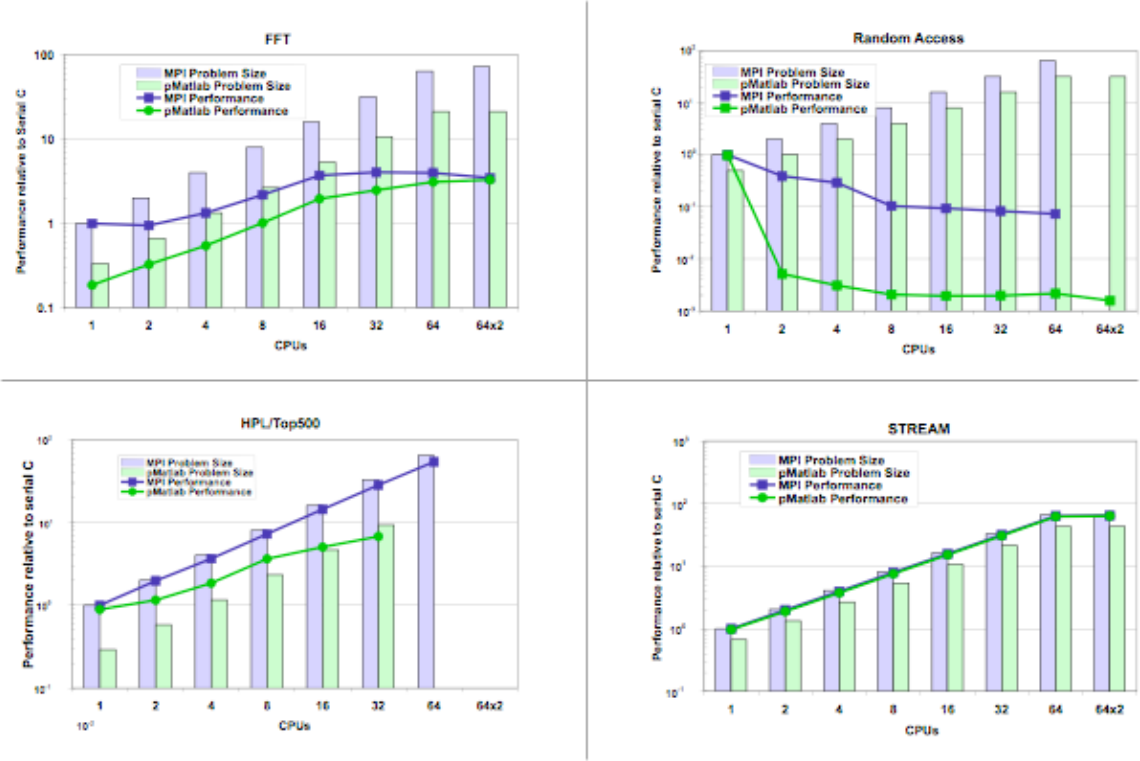

**Figure 14. pMatlab and C+MPI HPC Challenge Performance.** pMatlab can run problems that are typically ½ the size of C+MPI problem size. pMatlab performance varies from being comparable to the C+MPI code (FFT and STREAM), to somewhat slower (Top500), to a lot slower (RandomAccess). The figure presents performance relative to the 1 processor C+MPI case. The actual performance for the 1 processor C+MPI case can be found in Appendix A.

**Table 5a: Maximum problem size relative to the C+MPI single processor case on 128 processors.**

| Implementation | STREAM | FFT | RandomAccess | HPL(32) |
|---|---|---|---|---|
| C+MPI/C serial | 63.9 | 72.7 | 48 | 32.6 |
| pMatlab/C serial | 42.8 | 21.3 | 32 | 9.3 |
| C+MPI/pMatlab | 1.5 | 3.4 | 1.5 | 3.5 |

**Table 5b: Benchmark performance relative to the C+MPI single processor case on 128 processors.**

| Implementation | STREAM | FFT core* | RandomAccess | HPL(32) |
|---|---|---|---|---|
| C+MPI/C serial | 62.4 | 4.6 | $7.4 \times 10^{-2}$ | 28.2 |
| pMatlab/C serial | 63.4 | 4.3 | $1.6 \times 10^{-3}$ | 6.8 |
| C+MPI/pMatlab | 1 | 1 | 46 | 4 |

**Table 5c: Code Size Comparisons.** Code size is measured in terms of Source Lines Of Code (SLOC). The parallel code sizes of the HPC Challenge C+MPI reference code are taken from the HPC Challenge FAQ.

| Implementation | STREAM | FFT | RandomAccess | HPL |
|---|---|---|---|---|
| C+MPI | 347 | 787 | 938 | 8800 |
| pMatlab | 119 | 78* | 157 | 190 |
| C+MPI/pMatlab | 3 | 10 | 6 | 40 |

*Includes code used to create random waves, does not include code for initial and final all-to-all operations. Combined these should roughly offset each other.

## 4.1 STREAM

The STREAM benchmark consists of local operations on distributed vectors. The operations are copy, scale, add, and scale with addition defined as

$$a \leftarrow b + \alpha c$$

where a, b and c are double precision vectors of length $m$, with the constraint

$$\text{size}(a) + \text{size}(b) + \text{size}(c) = 24m \text{ bytes} > \tfrac{1}{4} \text{ system memory}$$

The goal of the benchmark is to measure local main memory bandwidth, so performance is reported in terms of bytes/sec

$$\text{Gigabytes/sec} = 10^{-9}\, 24\, m\, /\, \text{time}$$

The operations are embarrassingly parallel and are implemented entirely with the pMatlab fragmented PGAS approach (see Figure 6).

The max problem size of the pMatlab code is 1.5x smaller than the C+MPI code, which is due to the need to create intermediate temporary arrays. The need for temporaries are a side effect of most high level programming environments. The performance of the pMatlab code is the same as the C+MPI code. This is because the MATLAB interpreter recognizes the scale and add statement and replaces it with a call to the appropriate optimized Basic Linear Algebra Subroutine (BLAS). The pMatlab code is ~3x smaller than the C+MPI code due to the elimination of various for loops and the use of built in MATLAB functions.

## 4.2 FFT

The FFT benchmark performs a complex-to-complex 1D Fast Fourier Transform (FFT)

$$Z \leftarrow \text{FFT}(z)$$

Where Z and z are $m$ element double precision complex vectors, with the constraint

$$\text{size}(z+Z) = 32m \text{ bytes} > \tfrac{1}{4} \text{ system memory}$$

z input should be in linear "time" order. Z output should be in standard frequency order. Any necessary reordering time should be included. Regardless of how many actual operations are performed the performance in Gigaflops is reported using the standard radix 2 FFT algorithm operations count

$$\text{Gigaflops} = 10^{-9}\, 5\, m\, \log_2(m)\, /\, \text{time}$$

The standard parallel implementation of a 1D FFT performs two 2-D FFTs with a corner turn, or an all-to-all redistribution, between the two FFTs (see Figure 4). In our pMatlab implementation we deviated from the FFT specification in two ways. First, the input data is initialized using a random selection of cosine and sine waves, which does not affect performance, but is a significant aid to debugging the code. Second, our implementation uses an ordering scheme that eliminates initial and final all-to-all communication steps, which is more consistent with the use of this function for most real applications and provides a better predictor of 2D and 3D FFT performance. We have properly removed the time due to initial and final all-to-all steps in the C+MPI code so that a legitimate comparison can be made. The optimized pMatlab code (Figure 6) uses local arrays and the transpose_grid function with optimized message ordering previously discussed in section 2.2.

The max problem size of the pMatlab code is 3.5x smaller than the C+MPI code, which is due to the need to create intermediate temporary arrays. In addition, MATLAB internally uses a "split" representation for complex data types, while the serial FFTW library being called uses an "interleaved" representation. The result is that the data needs to be transformed between these representations which takes additional memory. On one processor the MATLAB FFT performance is ~5x slower than the C code, which is due to the time overhead required to perform the conversion between complex data storage formats. As the problem grows, the FFT time becomes dominated by the time to perform the all-to-all communication necessary between computation stages. Since these are primarily large messages, the performance of pMatlab becomes the same as the C+MPI code at large numbers of processors. The pMatlab code is ~10x smaller than the C+MPI code due to the use of a built in local FFT calls and the elimination of MPI messaging code.

### 4.3 RandomAccess

The RandomAccess benchmark generates a sequence of random array indices and uses these to update a large table. Let T be a table of size $2^m$ and let $\{a_i\}$ be a pseudo random stream of 64-bit integers of length $2^m+2$. Then for each $a_i$, we update the table as follows

$$T( AND(a_i,m-1) ) = XOR( T( AND(a_i,m-1), a_i)$$

with the additional constraints that each processor can buffer no more that 1024 updates and

$$size(Table) = 8m \text{ bytes} > \frac{1}{4} \text{ system memory}$$

The goal of the benchmark is to measure the rate at which atomic udates can be performed to global memory

$$\text{Giga Updates Per Second (GUPS)} = 10^{-9} \text{ NUPDATE/ time}$$

RandomAccess requires communication patterns that are significantly more complicated than STREAM or FFT. In addition, communication is sufficiently fine grained that there is significant overhead associated with computing global to local array indexing mappings every time a global array is accessed. Thus RandomAccess uses the pMatlab constructs to determine the global-to-local index mappings once, but then subsequently uses fragmented PGAS with direct message passing to perform the appropriate redistributions (see Figure 15). This methodology allows us to implicitly exploit the fact that the array redistributions are static. For example, each processor is able to compute in advance the optimal send order and optimal receive order of its messages so as to minimize contention. RandomAccess is a good illustration of how PGAS and messaging can work together to reduce the bookkeeping necessary for a parallel program, while still allowing a complex messaging scheme that is outside of the traditional PGAS formalism.

The max problem size of the pMatlab code is 1.5x smaller than the C+MPI code, which is due to the need to create intermediate temporary arrays. On one processor the pMatlab RandomAccess performance is comparable to the C+MPI code. However, on larger number of processors the pMatlab code is 45x slower than the C+MPI code. This performance difference is due to the large latency of using file I/O for communicating small messages (see section 3.4), which should be eliminated if pMatlab was built on a more traditional MPI implementation such as that used in DCT. The pMatlab code is 6x smaller than the C+MPI code.

```
Np = pMATLAB.comm_size;  % Set number of processors.
N = 2^20;                % Set dimensions of array.

Tmap = map([1 Np],{},0:Np-1);   % Create row map
% Create global unsigned int64 row vector.
Table = zeros(1,N,Tmap,'uint64');

Imy = global_block_range(X_table,2);  % Local index ranges
Iall = global_block_ranges(X_table,2); % All index ranges.

% Initialize table ...

% For each block of updates.

  ran = RandomAccess_rand(ran);  % Generate random indices.
  I = double(bitand(ran, TABLE_MASK)) + 1; % Compute table index.

  % Find indices for numbers that reside on rank i_cpu and send.
  for i_cpu = my_send_order
    j_cpu = find((I >= Iall(i_cpu+1,2)) & (I <= Iall(i_cpu+1,3)));
    MPI_Send(i_cpu, tag_number, pMATLAB.comm, ran(j_cpu));
  end

  % Concatenate receives from all processors.
  for i_cpu=my_recv_order
    ran_recv = [ran_recv MPI_Recv(i_cpu, tag_number, pMATLAB.comm)];
  end

% Compute local table index and perform vectorized XOR update.
Ilocal = double(bitand(ran_recv, TABLE_MASK)) - Imy(1) + 2;
Table(Ilocal) = bitxor(Table(Ilocal),ran_recv);
```

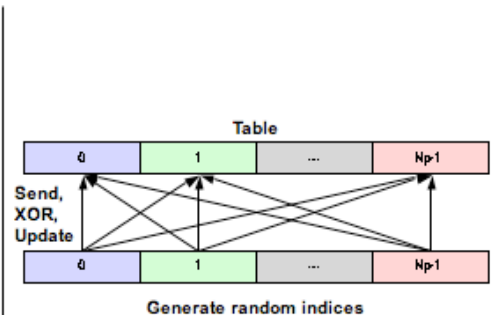

**Figure 15. RandomAccess Benchmark Code Highlights.** The first two lines set the various constants required by the program such as the number of processors and the size of the table. The next two lines create a map and a distributed table to be broken up equally among all the processors. The next two lines get the indices of the boundaries of the table which are used in the main loop to compute which indices to send to which processors using direct MatlabMPI messages.

**4.4 High Performance Linpack (Top500)**

The High-Performance Linpack (HPL) benchmark solves a dense linear system $Ax = b$. using LU factorization with partial pivoting, where b is an $n$ element vector, and A is an $n$x$n$ double precision matrix with the constraint

$$\text{size}(A) = 8n^2 \text{ bytes} > 1/2 \text{ system memory}$$

The LU factorization is the dominant computation step in this algorithm and is principally made up of repeated matrix multiplies. The traditional parallel algorithm uses a sophisticated 2D block-cyclic distribution for the matrix A. This algorithm has demonstrated very good performance even on computers with relatively slow networks. The pMatlab version uses a simpler, but poorer performing algorithm, using a 1D block distribution for A (see Figure 16). This algorithm and its theoretical performance limits are presented in detail in Appendix B. The pMatlab code uses distributed arrays to break up the array and keep track of the various global indices. A key step in the algorithm requires broadcasting the results to a subset of the other processors which is best done with a simple MPI multicast command.

The max problem size of the pMatlab code is 3.5x smaller than the C+MPI code, which is due to the need to create intermediate temporary arrays. In particular, the lower and upper triangular matrices are returned as full matrices, where in the C+MPI code these can be merged into a single array. The pMatlab code provides a 10x speedup on 32 processors, which is about 4x slower than the C+MPI code. The analysis in Appendix B shows that pMatlab is achieving the performance limits of the 1D block algorithm on the system. Improving the network of this hardware should significantly improve the pMatlab code performance, relative to the C+MPI code (see Figure B.1). The pMatlab code is 40x smaller than the C+MPI code. About 10x of this improvement is due to the higher-level abstractions from pMatlab and about 4x is due to using the simpler algorithm.

```
function [L,U,piv] = lu_parallel(A)
Np = pMATLAB.comm_size;  % Set number of processors.
my_index = pMATLAB.comm_rank + 1;  % Get processor index.
col_ranges = global_block_ranges(A,2);  % Cols belong to all processors.

% Get sizes and local part of A.
[m,n] = size(A);   Alocal = local(A);   nlocal = size(Alocal,2);

% ... initialize index counters ...

for p = 1:Np  % Loop over all processors.
  p_col = col_ranges(p,3) - col_ranges(p,2) + 1;  % Cols on processor p.
  if (p == my_index)  % Compute the LU of the p-th block.

    % ... compute index sets i and j ...

    [Alocal(i,j) pivp] = dgetrf(Alocal(i,j));  % Compute local LU and pivots.
    Lp = tril(Alocal(i,j));   % Get lower part.

    % Send Lp and pivp to all the higher processors and just pivp to lower.
    MPI_Mcast(p-1,p:(Np-1),tag_higher,comm,Lp,pivp);
    MPI_Mcast(p-1,0:(p-1),tag_lower,comm,pivp);

    % ... update index counters ...

  elseif (my_index > p)
    [Lp,pivp] = MPI_Recv(p-1,tag_higher,comm);  % Receive L and pivots
  elseif (my_index < p)
    pivp = MPI_Recv(p-1,tag_lower,comm);  % Receive pivots.
  end
  % ... apply pivots and weights ...
end
% ... Select L and U ...
```

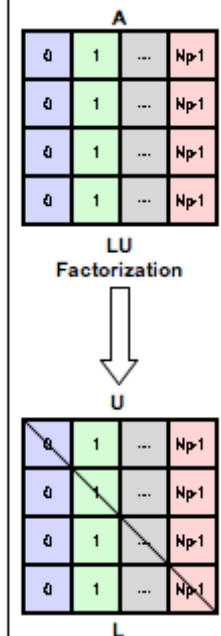

**Figure 16. HPL/Top500 Benchmark Code Highlights.** This algorithm uses the simpler 1D block distribution. The first four lines derive information about the parallel environment from the pMATLAB global variable and the input distributed matrix A. The core loop of the program performs a local solve of a rectangular LU, broadcasts the results to the remaining processors to then apply via a matrix multiply.

## 4.5 HPC Challenge Performance Summary

Returning to our initial metrics we see that relative to serial MATLAB all the pMatlab codes allow problems sizes to scale linearly with the number of processors. Likewise, they all experience significant performance improvements (with the exception of RandomAccess). Relative to C+MPI the pMatlab problem sizes are smaller by a factor of 2x and the performance of pMatlab on both the STREAM and FFT is comparable.

One approach to summarizing the performance of the HPC Challenge benchmarks is shown in Figure 17. The speedup and relative SLOC for each implementation were calculated with respect to a serial C/Fortran implementation. In this plot we see that with the exception of Random Access, the C+MPI implementations all fall into the upper-right quadrant of the graph, indicating that they deliver some level of parallel speedup, while requiring more SLOC than the serial code. As expected the serial MATLAB implementations do not deliver any speedup, but do all require fewer SLOC than the serial C/Fortran code. The pMatlab implementations (except Random Access) fall into the upper-left quadrant of the graph, delivering parallel speedup while requiring fewer lines-of-code.

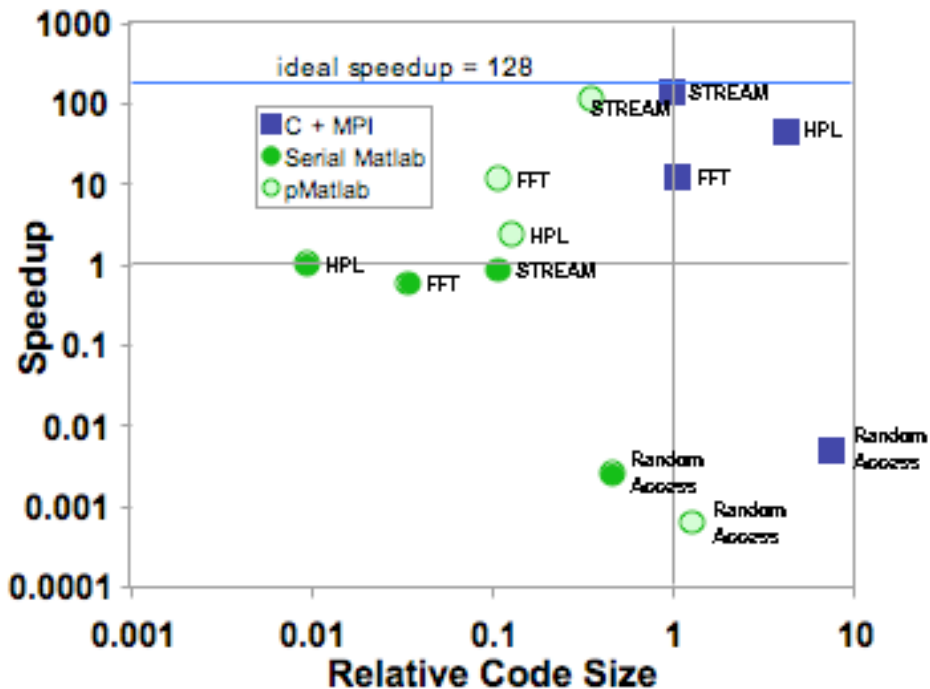

**Figure 17. Speedup vs Code Size.** Speedup (relative to serial C) vs code size (relative to serial C). The upper right quadrant is the traditional HPC regime: more coding is required to give more performance and most of the C+MPI codes fall here. The lower left quadrant is the traditional regime of serial high level languages that produce much smaller codes, but are slower. RandomAccess lies in the lower right and represents algorithms that are simply a poor match to the underlying hardware. The upper left quadrant is where most of the pMatlab implementations are found and represent smaller codes that are delivering some speedup.

## 5. User Results

The true measure of a success for any technology is its effectiveness for real users. Table 6 highlights several projects that are using pMatlab on the MIT Lincoln Laboratory interactive

LLGrid system [Reuther2004]. The projects are drawn from the approximately 100 current users and are representative of the user base. Of particular interest are the columns showing the time to parallelize and what parallelization enables. The time to parallelize shows both how quickly MATLAB code can be converted from serial code to parallel code as well as how quickly the user is able to get the parallel code running on the LLGrid compute facility. The applications that parallelization enables include scenarios in which larger data sets, more thorough parameter set exploration, and more complex simulations can be considered.

Nearly all of these applications involve embarrassingly parallel problems most similar to the STREAM type of problem. Interestingly, because MATLAB is such an array oriented language users find PGAS a very natural way to express embarrassingly parallel problems. For these types of applications the coding overhead is much smaller than message passing. In addition, PGAS naturally decomposes problems into their largest natural units, which maximizes the local performance. In contrast a client/server approach tends to decompose problems into their smallest functional units and incur a higher overhead.

**Table 6. Selected pMatlab Applications.** The first and last columns provide a brief description of the code and what the parallel version of the code has enabled. The middle column shows estimated time to write the original serial code and the additional time to parallelize the code with pMatlab and get it running well on the LLGrid system.

| Code Description | Serial / Parallel Dev Time (hours) | Parallelization Enables More or Faster |
|---|---|---|
| Missile & Sensor Simulations | 2000 / 8 | Higher fidelity radar |
| First-principles LADAR | 1300 / 1 | Speckle image simulations |
| Analytic TOM Leakage | 40 / 0.4 | Parameter space studies |
| Hercules Metric TOM | 900 / 0.75 | Monte Carlos |
| Coherent laser propagation | 40 / 1 | Run time |
| Polynomial coefficient approx. | 700 / 8 | Faster training algorithm |
| Ground motion tracker | 600 / 3 | Faster & larger data sets |
| Automatic target recognition | 650 / 40 | Target classes & scenarios |
| Hyper-spectral Image Analysis | 960 / 6 | Larger datasets of images |

## 6. Conclusions

pMatlab is a unique high performance, high productivity parallel MATLAB library. It combines the productivity inherent in the MATLAB programming language with global array semantics, allowing MATLAB users to exploit distributed systems with only minor changes to the code. The underlying communication layer, MatlabMPI, is comparable in performance to C+MPI for large message sizes. Introduction of maps for numerical arrays allows for separation of functional programming from mapping the program to a parallel architecture. The implementation is small (~3,000 lines of code). The implementation of the HPC Challenge benchmark suite using the pMatlab library allows for comparison with equivalent C+MPI codes. These results indicate that pMatlab can achieve comparable performance to C+MPI at usually $1/10^{th}$ the code size. Finally, implementation data collected from a sample of 10 real pMatlab applications indicate that users are typically able to go from a serial code to a well-performing pMatlab code in about 3 hours while changing less than 1% of their code.

**Acknowledgements**

The authors would like to thank a number of individuals who have contributed to this work: Bob Bond for his vision and insight throughout this project; Hahn Kim for his work on the pMatlab library and benchmarking; Andy Funk for his benchmarking analysis; Albert Reuther for leading the LLgrid project and providing us the pMatlab user analysis; Cleve Moler and Ryan Haney for their assistance with the parallel LU algorithm; and Charlie Rader for his assistance with the parallel FFT algorithm. Finally, we would like to thank Ken Senne, Dave Martinez, John Grosh, and Robert Graybill for supporting this project.**References**
[Cardelli1985] Cardelli, L., Wegner, P. On Understanding Types, Data Abstraction, and Polymorphism. *ACM Computing Surveys* 17(4), 1985.
[Choy2003] Choy, R. 2003. Parallel Matlab survey. http://supertech.lcs.mit.edu/~cly/survey.html
[Choy2005] Choy, R., Edelman A. 2005. Parallel MATLAB: doing it right. *Proceedings of the IEEE* 93(2).
[CMTM] Cornell Multitask Toolbox for MATLAB (CMTM), http://www.cs.cornell.edu/Info/People/lnt/multimatlab.html
[Cummings1998] J. C. Cummings, J. A. Crotinger, S. W. Haney, W. F. Humphrey, S. R. Karmesin, J. V. Reynders, S. A. Smith, and T. J. Williams. Rapid application development and enhanced code interoperability using the POOMA framework. In *Proceedings of the SIAM workshop on Object-oriented methods and code interoperability in scientific and engineering computing (OO98)*, Oct. 1998
[Dean2005] Dean, L., Grad-Freilich, S., Kepner, J., Reuther, A. Distributed and Parallel Computing with MATLAB. Tutorial presented at Supercomputing 2005, Nov 12, Seattle, WA
[DeLuca1997] DeLuca, C. M., Heisey, C. W., Bond, R. A., Daly, J. M. A portable object-based parallel library and layered framework for real-time radar signal processing. In *Proc. 1$^{st}$ Conf. International Scientific Computing in Object-Oriented Parallel Environments (ISCOPE '97)*, Pages: 241-248.
[Dongarra1994] Dongarra, J., van de Geijn, R., Walker, D. Scalability Issues Affecting the Design of a Dense Linear Algebra Library. *Journal of Parallel and Distributed Computing*, Volume 22, Pages 523-537, 1994
[El-Ghazawi2005] El-Ghazawi, T., Carlson, W., Sterling, T., Yelick, K. UPC: Distributed Shared Memory Programming, Published by John Wiley and Sons- May, 2005
[Falcon] Falcon Project: Fast Array Language Computation, http://www.csrd.uiuc.edu/falcon/falcon.html
[Funk2005] Funk, A., Kepner, J., Basili, V., Hochstein, L. A Relative Development Time Productivity Metric for HPC Systems. *Proceedings of the High Performance Embedded Computing Workshop (HPEC2005)*, Lexington, MA, September 20-22, 2005.
[Johnson2004] Johnson, P. (editor), *Proceedings of 26$^{th}$ International Conference on Software Engineering (ICSE 2004), Edinburgh, Scontland, UK*, May 23-28.
[Johnson2005] Johnson, P. (editor), *Proceedings of 27$^{th}$ International Conference on Software Engineering (ICSE 2005), St. Louis, Missouri,* May 15-21.
[Kepner2004a] Kepner, J., Ahalt, S. 2004. MatlabMPI. *Journal of Parallel and Distributed Computing*, 2004, Volume 64, Issue 8, Pages: 997 - 1005
[Kepner2004b] Kepner, J (editor). 2004. Special issue on HPC Productivity, *International Journal of High Performance Computing Applications* 18(4).
[Koelbel1994] Koelbel, C., The High performance Fortran handbook, MIT Press, 199427

**Appendix A: Benchmark System**

All the performance data collected in this paper were obtained using the LLGrid system at MIT Lincoln Laboratory [Reuther2004]. The system consists of ~150 nodes connected by Gigabit Ethernet. Each node has 2 Gigabit Ethernet interfaces: one gigabit interface to the Lincoln Laboratory LAN (LLAN), and one gigabit inter-cluster interface. The network switches are connected directly to the LLAN backbone via fiber. Furthermore to enhance the communication of the file I/O based communication system each node mounts the local disk drive of all the other nodes. Each node is configured as follows:

    Processors: Dual 3.2 GHz EM-64T Xeon (P4)
    Bus: 800 MHz front-side bus
    Memory: 6 Gigabyte RAM
    Disk: Two 144 GB SCSI hard drives

Main Network: Two Gig-E Intel interfaces
Management Network: 10/100 Ethernet interface
Operating System:  Red Hat Linux ES 3

Table A.1 provides the actual benchmark values of the HPC Challenge benchmark suite on the LLGrid system for the C+MPI single processor case.

**Table A.1: C+MPI single processor HPC Challenge.** Maximum problem size and performance of the HPC Challenge benchmarks for C+MPI implementation on a single processor.

| Benchmark | Maximum Problem Size (GB) | Performance |
|---|---|---|
| STREAM | 4.6 | 2.79 GB/s |
| FFT | 3 | .43 GFlops |
| RandomAccess | 4 | $2.3 \times 10^{-3}$ GUPS |
| HPL/Top500 | 4.3 | 4.04 GFlops |

**Appendix B: 1D LU Algorithm Performance Analysis**

This appendix presents an analysis of the theoretical performance achievable for a LU factorization using only a 1D block distribution.  Extensive analysis of LU performance has been performed for optimal 2D block cyclic distributions [Dongarra2004], which shows the excellent scalability of this distribution.  However, more recently it is become apparent that there is a complexity performance trade off associated with using 2D block cyclic distributions (i.e. they are a lot harder to program) and so it is worth examining the performance of a 1D block distribution (like the one used in for the pMatlab implementation of Top500) so we have a clear understanding of the performance.

First we define the time for an ideal NxN LU factorization on P processors

$$T_N^{ideal}(P) = \tfrac{2}{3} N^3 t_{calc} / P$$

where $t_{calc}$ is the time for one floating point operation on one processor.  Furthermore let $r = N/P$, so that in terms of P and r the ideal performance is

$$T_r^{ideal}(P) = \tfrac{2}{3} P^2 r^3 t_{calc}$$

We will further restrict ourselves to scaled problems such that the problem size grows linearly with the number or processors.  In this case, we have the additional constraint

$$r N = c_{mem} \qquad \text{or} \qquad r = \sqrt{\tfrac{c_{mem}}{P}}$$

where  $c_{mem}$ is the number of 8 byte double precision numbers that will fit on one processor.

Now let's consider the time to perform a parallel LU factorization using a 1D block distribution.  The algorithm consists of k=1,…,P steps and at each step three operations must be completed before the next step can begin.

First, a local LU factorization of a (N - (k-1)r) x r matrix is performed on the $k^{th}$ processor

$$\tfrac{2}{3}(N - (k-1)r)\,r^2\,t_{calc}$$

Second, the result of this local LU factorization is broadcast from the $k^{th}$ processors to P-k processors

$$B(P-k)\,(N - (k-1)r)\,r\,t_{comm}$$

where $t_{comm}$ is the average time to send 8 bytes between two processors (assuming large messages), and B(P) is the "broadcast" parameter which is the penalty associated with sending the same message to many processors. In an ideal broadcast B(P) = 1 and a processor can send to many processors in the same time it takes to send to one. In the worst case B(P) = P. Most networks are somewhere in between and B(P) = $\log_2(P)$ is typical. The third and final step is to apply the local LU factorization to the local part of the matrix stored on the processor using a matrix matrix multiply operation

$$(N - (k-1)r)\,r^2\,t_{calc}$$

From this point forward we will combine the formulas for the first and third steps since they only differ by a constant.

Next we sum the above steps over k=1,…,P and reformulate in terms of P and r, which yields the total time required to compute the LU and the matrix multiply

$$\sum_{k=1}^{P} \tfrac{5}{3}(N - (k-1)r)r^2 t_{calc} = \tfrac{5}{6}P^2 r^3 t_{calc}\left[1 + P^{-1}\right]$$

where we have used the summation identity

$$\sum_{k=1}^{P} k = \tfrac{1}{2}P(P+1)$$

Similarly, we sum up the broadcast term for two cases, B(P) = 1

$$\sum_{k=1}^{P} (N - (k-1)r)\,r\,t_{comm} = \tfrac{1}{2}P^2 r^2 t_{comm}\left[1 + P^{-1}\right]$$

and B(P) = P

$$\sum_{k=1}^{P} (P-k)\,(N - (k-1)r)\,r\,t_{comm} = \tfrac{1}{3}P^2 r^2\,t_{comm}\left[P - P^{-1}\right]$$

where we have used the summation identity

$$\sum_{k=1}^{P} k^2 = \tfrac{1}{3}P^3 + \tfrac{1}{2}P^2 + \tfrac{1}{6}P$$

Finally, for each case, we sum the computation and network terms and normalize to the ideal time. In the case of an ideal network, B(P) = 1, the ratio of the 1D panel algorithm to the ideal time is

$$\tfrac{5}{4}\left[1+P^{-1}\right] + \tfrac{3}{4}\sqrt{\tfrac{P}{c_{mem}}}(t_{comm}/t_{calc})\left[1+P^{-1}\right]$$

Interestingly, the first computation term goes to a constant value of 5/4 for large P, which indicates the 1D block algorithm is always at least 25% less efficient due to load imbalance. In addition, the second communication term will grow with the square root of P for large P. However, for typical values of $c_{mem}$ (~$2^{29}$) on an high performance system with a fast network ($t_{comm}/t_{calc}$ ~ 8) the constant in front the communication term is quite small and the overhead due to the broadcast doesn't become significant until P > 100,000. Thus, for such a system the 1D block distribution will scale well.

In the worst case network, B(P) = P, the ratio to the ideal algorithm is

$$\tfrac{5}{4}\left[1+P^{-1}\right] + \tfrac{1}{2}\sqrt{\tfrac{P}{c_{mem}}}(t_{comm}/t_{calc})\left[P-P^{-1}\right]$$

which is nearly the same as the best case formula except that the broadcast term has an additional factor P which causes it to become significant for much lower values of P. The system used to obtain the results in the paper (see Appendix A) is best approximated by this model with values $c_{mem}$ ~ ~$2^{27}$ and $t_{comm}/t_{calc}$ ~ 400. The measured and predicted values are shown in Figure B.1, along with the predicted values for a network that is 10x faster as well as network with B(P) = 1. As we can see, the pMatlab code is achieving the performance limit presented by the algorithm and will move closer to the more efficient (and more complex) 2D block cyclic algorithm as the network hardware is improved.

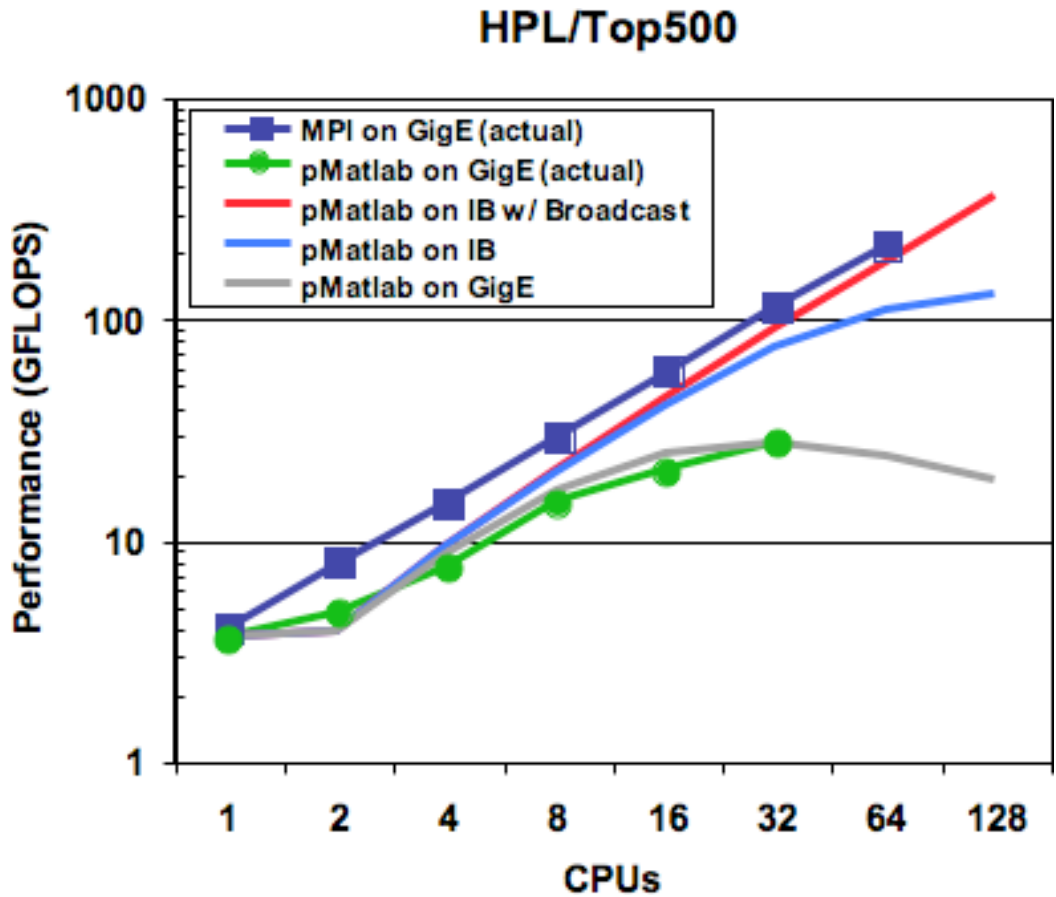

**Figure B.1. Measured and predicted LU performance.** The LU performance model is shown for several different configurations of Gigabit Ethernet (GigE) and InfiniBand (IB) networks. The lowest performing network agrees well with the pMatlab data. As the network is improved the performance should approach that of the C+MPI code, which uses the more efficient (and more complex) 2D block cyclic distribution.